\documentclass{bioinfo}
\copyrightyear{2013}
\pubyear{2013}

\usepackage{natbib}
\begin{document}
\firstpage{1}

\title{Aligning sequence reads, clone sequences and assembly contigs with BWA-MEM}

\author[Li]{Heng Li}

\address{Broad Institute of Harvard and MIT, 7 Cambridge Center, Cambridge, MA 02142, USA}

\history{Received on XXXXX; revised on XXXXX; accepted on XXXXX}
\editor{Associate Editor: XXXXXXX}
\maketitle

\begin{abstract}
\section{Summary:} BWA-MEM is a new alignment algorithm for aligning sequence
reads or assembly contigs against a large reference genome such as human.
It automatically chooses between local and end-to-end alignments, supports
paired-end reads and performs chimeric alignment. The algorithm is robust to
sequencing errors and applicable to a wide range of sequence lengths from 70bp
to a few megabases. For mapping 100bp sequences, BWA-MEM shows better
performance than several state-of-art read aligners to date.

\section{Availability and implementation:} BWA-MEM is implemented as a
component of BWA, which is available at http://github.com/lh3/bwa.

\section{Contact:} hengli@broadinstitute.org
\end{abstract}

\section{Introduction}

Most short-reads mappers for next-generation sequencing (NGS) data were
developed when sequence reads were about 36bp in length. For 36bp reads, it is
reasonable to require end-to-end alignment (i.e. every read base to be aligned
to the reference) and to only report hits within certain hamming or edit
distance.  However, with emerging technologies and improved chemistry, NGS
reads are not short any more, which poses new challenges to read alignment. For
100bp or longer reads, it becomes more important to allow long gaps under the affine-gap
penalty and to report multiple non-overlapping local hits potentially caused by
structural variations or misassemblies in the reference genome. Many short-read
alignment algorithms are not applicable or not preferred for mapping longer
reads. At the same time, although several mature algorithms exist for aligning
capillary sequence reads, they are slow and lack features for analyzing
large-scale NGS data. Fast moving NGS technologies keep pressing for the
development of new alignment algorithms.

In this background, a few long-read alignment algorithms, notably including
BWA-SW~\citep{Li:2010fk}, Bowtie2 \citep{Langmead:2012fk},
Cushaw2~\citep{Liu:2012uq} and GEM~\citep{Marco-Sola:2012kx}, have been
developed. However, they all have some weakness. BWA-SW is slower than bowtie2
for 100bp reads at a comparable accuracy and less accurate then Cushaw2 at a
comparable speed. Bowtie2 and Cushaw2 are slower for 600bp reads (see Results).
While GEM is both fast and accurate for up to approximately 1000bp reads, it
mandates end-to-end alignment and does not perform affine-gap alignment, which
limits its uses for long-read alignment. Even for typical sequence reads ranged
from 100bp to 1000bp in length, no mappers work well across the full spectrum.
At the same time, the increasing read length not only calls for new alignment
algorithms, but also opens the opportunity to de novo assembly which typically
results in contigs ranged from several hundred base pairs to a few megabases.
Very few algorithms are able to align sequences of such variable lengths
at high accuracy and to be robust to translocations in the assembly. All these
concerns motivated us to explore a new alignment algorithm.

\begin{methods}

\section{Methods}

\subsection{Aligning a single query sequence}

\subsubsection{Seeding and re-seeding} BWA-MEM follows the canonical
seed-and-extend paradigm. It initially seeds an alignment with supermaximal
exact matches (SMEMs) using an algorithm we found previously~\citep[Algorithm
5]{Li:2012fk}, which essentially finds at each query position the longest exact
match covering the position. However, occasionally the true alignment may not
contain any SMEMs.  To reduce mismappings caused by missing seeds, we introduce
re-seeding. Suppose we have a SMEM of length $l$ with $k$ occurrences in the
reference genome. If $l$ is too large (over 28bp by default), we re-seed
with the longest exact matches that cover the middle base of the SMEM and occur
at least $k+1$ times in the genome. Such seeds can be found by requiring a
minimum occurrence in the original SMEM algorithm.

\subsubsection{Chaining and chain filtering} We call a group of seeds that are
colinear and close to each other as a \emph{chain}. We greedily chain the seeds
while seeding and then filter out short chains that are largely contained in a
long chain and are much worse than the long chain (by default, both 50\% and
38bp shorter than the long chain). Chain filtering aims to reduce unsuccessful
seed extension at a later step. Each chain may not always correspond to a final
hit. Chains detected here do not need to be accurate.

\subsubsection{Seed extension} We rank a seed by the length of the chain it
belongs to and then by the seed length. For each seed in the ranked list, from
the best to the worst, we drop the seed if it is already contained in an
alignment found before, or extend the seed with a banded affine-gap-penalty
dynamic programming (DP) if it potentially leads to a new alignment.

BWA-MEM's seed extension differs from the standard seed extension in two
aspects. Firstly, suppose at a certain extension step we come to reference
position $x$ with the best extension score achieved at query position $y$.
BWA-MEM will stop extension if the difference between the best extension score
and the score at $(x,y)$ is larger than $Z+|x-y|\times p_{\rm gapExt}$, where
$p_{\rm gapExt}$ is the gap extension penalty and $Z$ is an arbitrary cutoff.
This heuristic avoids extension through a poorly aligned region with good
flanking alignment. It is similar to the X-dropoff heuristic in
BLAST~\citep{Altschul:1997vn}, but does not penalize long gaps in one of the
reference or the query sequences.

Secondly, while extending a seed, BWA-MEM tries to keep track of the best
extension score reaching the end of the query sequence. If the difference
between the best score reaching the end and the best local alignment score is
below a threshold, the local alignment will be rejected even if it has a higher
score. BWA-MEM uses this strategy to automatically choose between local and
end-to-end alignments. It may align through true variations towards the end of
a read and thus reduces reference bias, while avoids introducing excessive
mismatches and gaps which may happen when we force a chimeric read into an
end-to-end alignment.

\begin{figure}[tb]
\centering
\includegraphics[width=0.5\textwidth]{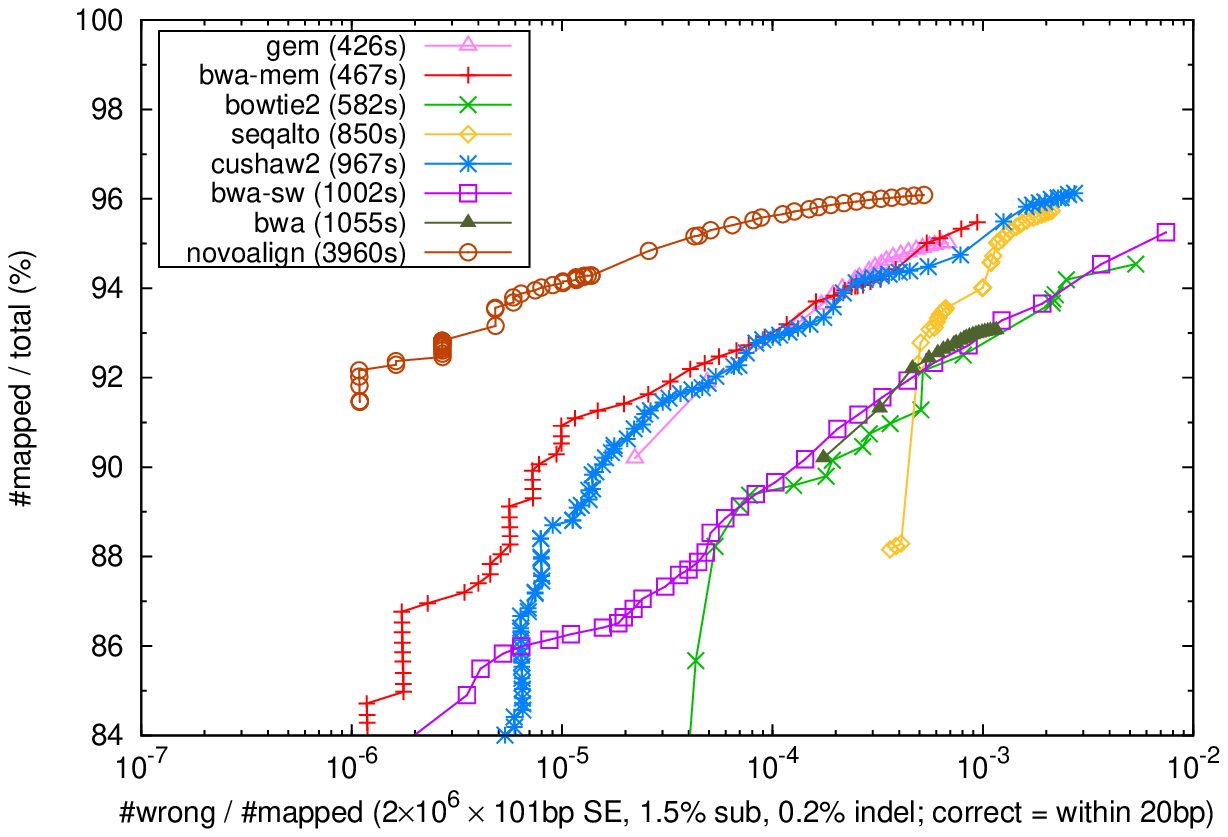}\\
\includegraphics[width=0.5\textwidth]{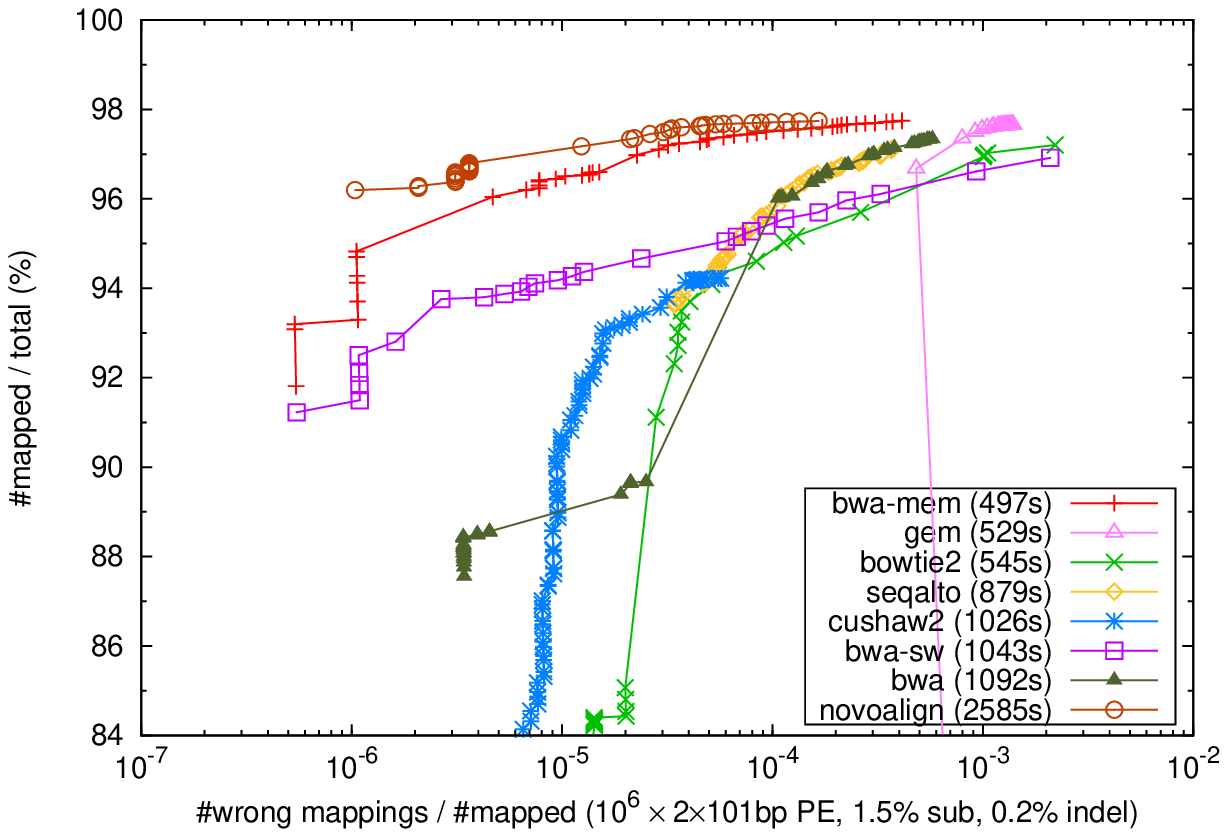}
\caption{Percent mapped reads as a function of the false alignment rate under
different mapping quality cutoff. Alignments with mapping quality 3 or lower
are excluded. An alignment is \emph{wrong} if after correcting clipping, its
start position is within 20bp from the simulated position. $10^6$ pairs
of 101bp reads are simulated from the human reference genome using wgsim
(http://bit.ly/wgsim2) with 1.5\% substitution errors and 0.2\% indel variants.
The insert size follows a normal distribution $N(500,50^2)$. The reads are
aligned back to the genome either as single end (SE; top panel) or as paired
end (PE; bottom panel). GEM is configured to allow up to 5 gaps and to output
suboptimal alignments (option `--e5 --m5 --s1' for SE and `--e5 --m5 --s1 --pb'
for PE). GEM does not compute mapping quality. Its mapping quality is estimated
with a BWA-like algorithm with suboptimal alignments available. Other mappers
are run with the default setting except for specifying the insert size
distribution.
The run time in seconds on a single CPU core is shown in the
parentheses.}\label{fig:eval}
\end{figure}

\subsection{Paired-end mapping}

\subsubsection{Rescuing missing hits}
Like BWA~\citep{Li:2009uq}, BWA-MEM processes a batch of reads at a time.
For each batch, it estimates the mean $\mu$ and the variance $\sigma^2$
of the insert size distribution from reliable single-end hits. For the
top 100 hits (by default) of either end, if the mate is unmapped in
a window $[\mu-4\sigma,\mu+4\sigma]$ from each hit, BWA-MEM performs SSE2-based
Smith-Waterman alignment (SW; \citealt{Farrar:2007hs}) for the mate within the
window. The second best SW score is recorded to detect potential mismapping in
a long tandem repeat. Hits found from both the single-sequence alignment and SW
rescuing will be used for pairing. 

\subsubsection{Pairing} Given the $i$-th hit for the first read and $j$-th hit
for the second, BWA-MEM computes their distance $d_{ij}$ if the two hits are in
the right orientation, or sets $d_{ij}$ to infinity otherwise. It scores the
pair $(i,j)$ with $S_{ij}=S_i+S_j-\min\{-a\log_4 P(d_{ij}),U\}$, where $S_i$
and $S_j$ are the SW score of the two hits, respectively, $a$ is the matching
score and $P(d)$ gives the probability of observing an insert size larger than
$d$ assuming a normal distribution. `$\log_4$' arises when we interpret SW
score as odds ratio~\citep{Durbin:1998uq}. $U$ is a threshold that controls
pairing: if $d_{ij}$ is small enough such that $-a\log_4 P(d_{ij})<U$, BWA-MEM
prefers to pair the two ends; otherwise it prefers the unpaired alignments.
BWA-MEM chooses the pair $(i,j)$ that maximizes $S_{ij}$ as the final
alignments for both ends.  This pairing strategy jointly considers the
alignment scores, the insert size and the possibility of chimeric read pairs.

\end{methods}

\section{Results and Discussions}

We evaluated the performance of BWA-MEM on simulated data together with
NovoAlign (http://novocraft.com), GEM, Bowtie2, Cushaw2, SeqAlto~\citep{Mu:2012fk}, 
BWA-SW
and BWA (Figure~\ref{fig:eval}). On accuracy, NovoAlign is the best. BWA-MEM
is close to NovoAlign for PE reads and is comparable to GEM and Cushaw2 for SE.
On speed, BWA-MEM is similar to GEM and Bowtie2 for this data set, but is about
6 times as fast as Bowtie2 and Cushaw2 for a 650bp long-read data set.

We speculate BWA-MEM is more performant for longer reads firstly because of its
advanced seeding algorithm, which identifies most standing seeds with one pass
of the read, and secondly because of banded dynamic programming, which
guarantees linear time complexity in the length of query sequences. BWA-MEM and
BWA-SW are also able to identify chimeric reads, a crucial feature for contig
alignment but lacked in most NGS long-read mappers. Our earlier
paper~\citep{Li:2010fk} discussed this in details.

To evaluate the performance for even longer query sequences, we aligned the
{\it E. coli} K-12 MG1655 substrain (AC:NC\_000913; 4.6Mb in length) against
the 536 strain (AC:NC\_008253) with both BWA-MEM and
nucmer~\citep{Kurtz:2004zr}. Nucmer finished the alignment in 25 seconds,
giving 105,505 substitution differences between the strains. BWA-MEM is slower.
It finished the alignment in 131 seconds, reporting 104,321 substitutions of
which 102,241 overlapping with the nucmer output. Manual examination of the
substitutions unique to each aligner reveals that most of them fall in short
regions with very high divergence. It is unclear which aligner is better. Note
that although nucmer is faster in the evaluation, only BWA-MEM scales well to
large genomes.

BWA-MEM is a fast and accurate aligner for sequence reads and is one of the few
that work well for both 70bp reads and long sequences up to a few megabases.
Technically, it is possible to make BWA-MEM even faster for long sequences by
using SSE2-based banded DP and by restricting DP to regions not covered by a
long exact match. Seeding is the bottleneck for short sequences, while banded
DP is the bottleneck for long sequences.

\section{Acknowledgements}
\paragraph{Funding\textcolon} NIH 1U01HG005208-01.
\bibliography{bwamem}

\begin{thebibliography}{}

\bibitem[Altschul et~al., 1997]{Altschul:1997vn}
Altschul, S.~F. et~al. (1997).
\newblock {Gapped BLAST and PSI-BLAST: a new generation of protein database
  search programs}.
\newblock {\em Nucleic Acids Res}, 25:3389--402.

\bibitem[Durbin et~al., 1998]{Durbin:1998uq}
Durbin, R. et~al. (1998).
\newblock {\em Biological sequence analysis}.
\newblock Cambridge University Press.

\bibitem[Farrar, 2007]{Farrar:2007hs}
Farrar, M. (2007).
\newblock Striped smith-waterman speeds database searches six times over other
  simd implementations.
\newblock {\em Bioinformatics}, 23:156--61.

\bibitem[Kurtz et~al., 2004]{Kurtz:2004zr}
Kurtz, S. et~al. (2004).
\newblock Versatile and open software for comparing large genomes.
\newblock {\em Genome Biol}, 5:R12.

\bibitem[Langmead and Salzberg, 2012]{Langmead:2012fk}
Langmead, B. and Salzberg, S.~L. (2012).
\newblock {Fast gapped-read alignment with Bowtie 2}.
\newblock {\em Nat Methods}, 9:357--9.

\bibitem[Li, 2012]{Li:2012fk}
Li, H. (2012).
\newblock {Exploring single-sample SNP and INDEL calling with whole-genome de
  novo assembly}.
\newblock {\em Bioinformatics}, 28:1838--44.

\bibitem[Li and Durbin, 2009]{Li:2009uq}
Li, H. and Durbin, R. (2009).
\newblock Fast and accurate short read alignment with burrows-wheeler
  transform.
\newblock {\em Bioinformatics}, 25:1754--60.

\bibitem[Li and Durbin, 2010]{Li:2010fk}
Li, H. and Durbin, R. (2010).
\newblock Fast and accurate long-read alignment with burrows-wheeler transform.
\newblock {\em Bioinformatics}, 26:589--95.

\bibitem[Liu and Schmidt, 2012]{Liu:2012uq}
Liu, Y. and Schmidt, B. (2012).
\newblock Long read alignment based on maximal exact match seeds.
\newblock {\em Bioinformatics}, 28:i318--i324.

\bibitem[Marco-Sola et~al., 2012]{Marco-Sola:2012kx}
Marco-Sola, S. et~al. (2012).
\newblock {The GEM mapper: fast, accurate and versatile alignment by
  filtration}.
\newblock {\em Nat Methods}, 9:1185--8.

\bibitem[Mu et~al., 2012]{Mu:2012fk}
Mu, J.~C. et~al. (2012).
\newblock Fast and accurate read alignment for resequencing.
\newblock {\em Bioinformatics}, 28:2366--73.

\end{thebibliography}

\end{document}